\documentclass[11pt,twoside]{article}
\usepackage{asp2010}

\resetcounters

\begin{document}

\aspvoltitle{ADASS XXI}
\aspvolume{TBD}
\aspcpryear{2012}
\aspvolauthor{P. Ballester and D. Egret, eds.}

\title{What is a Spectrum?}
\author{Adam S. Bolton$^1$, Stephen Bailey$^2$, Joel Brownstein$^1$,
Parul Pandey$^1$, David Schlegel$^2$, Yiping Shu$^1$
\affil{$^1$Department of Physics and Astronomy, The University of Utah}
\affil{$^2$Physics Division, Lawrence Berkeley National Lab}}

\begin{abstract}
This contribution describes the ``spectro-perfectionism'' algorithm
of Bolton \& Schlegel (2010)
that is being implemented within the
Baryon Oscillation Spectroscopic Survey (BOSS) of the Sloan
Digital Sky Survey III (SDSS-III), in terms of its
potential to deliver
Poisson-limited sky subtraction and lossless compression of the input
spectrum likelihood functional given raw CCD data.
\end{abstract}

The Baryon Oscillation Spectroscopic Survey (BOSS) of the Sloan Digital
Sky Survey III \citep[SDSS-III][]{2011AJ....142...72E}
is the largest extragalactic spectroscopic
survey to date, both in the number of spectra being collected and in
the volume of the universe being mapped. BOSS has been delivering
survey-quality spectra since December 2009, and is on track to obtain
spectra of more than 1.5 million galaxies by mid-2014.  While BOSS
target galaxies are fainter than the galaxies observed in the original
SDSS, the night-sky foreground is as bright as ever. This decreasing
signal-to-foreground trend---which will only be exacerbated in future
redshift surveys such as the proposed BigBOSS
experiment \citep{2011arXiv1106.1706S}---demands a
new approach to spectroscopic data analysis.  This contribution
to the proceedings describes the ``spectro-perfectionism''
algorithm of \citep[][hereafter B\&S]{2010PASP..122..248B}
in terms of its potential to deliver two important
criteria for faint-object spectroscopy: (1) Poisson-limited
night-sky foreground subtraction, and (2) lossless likelihood
functional compression.  The B\&S algorithm is based upon
extraction via forward-modeling to the raw CCD data using the
full two-dimensional (2D) point-spread function (PSF) of
the spectrograph, and is currently in the software implementation
phase for eventual deployment within the BOSS survey.

The current state of the art for extraction of spectra
in optical astronomy is the \textit{optimal extraction} method described
by \citet{1985MNRAS.213..971H} and \citet{1986PASP...98..609H},
which models the spectrum in each successive
CCD pixel row (corresponding to a particular effective
wavelength) via a maximum-likelihood scaling of a spatial
cross-sectional profile.
The shortcomings of this method can best be
understood by considering how it extracts an emission line,
such as one of the OH features that are prominent redward of
7000\,\AA\@.  If the PSF is a non-separable function of $x$ and $y$
in CCD coordinates (such as will be the case for a PSF consisting of
a core plus wings, or a PSF with significant tilted ellipticity and/or skewness),
then the actual spatial cross-sectional profile will depend upon the
location of the cross section within the spectrum (e.g., a cross section
through the wing will be broader than a cross section through the core).
By extracting with a single mean profile $p(x)$ rather than the
true cross-sectional profile $q(x)$ in a given row,
a fractional bias in the extracted
counts is introduced which is given by
\begin{equation}
b = 1 - \left[ \int dx \, p(x) q(x) \right] \left[ \int dx \, q^2(x) \right]^{-1} ~.
\end{equation}
In practice this bias may be small, but when coupled to variations
in the spectrograph PSF across the field of view of a multifiber spectrograph,
it can result in significant systematic residuals after the subtraction
of a model sky spectrum.  For very faint galaxy spectroscopy, we wish
to push these residuals well below the 1\% level, and hence we are
pursuing extraction using the mathematically correct forward
model given by the 2D PSF rather than the mean 1D cross-section.
Given a sufficiently accurate model for the PSF and its variation across
the instrument, we can construct a noise-limited model for the raw CCD data,
and can furthermore model the common sky spectrum \textit{upstream} from the variation
of the PSF.

To realize the full benefit of 2D PSF-based extraction, we
develop a mathematically explicit answer to the question:
what is a spectrum?
Let the vector $\mathbf{f}$ represent an extracted 1D spectrum
in the sense commonly understood in astronomy.
Its most important function
is to permit quantitative inference about the object under
observation.  For this purpose, the full
information content of the spectrum
is not only in the vector $\mathbf{f}$, but also in the
estimate of its noise properties and its resolution.
We will represent the noise by the statistical
covariance matrix $\mathbf{C}$ of the pixels in the spectrum,
and the resolution by a matrix $\mathbf{R}$ that encodes
the ``line-spread function'' (LSF) of the spectrum.
If it is diagonal, $\mathbf{C}$ can be represented by a
vector of pixel errors (and this is often assumed even when
it is not true).  The matrix $\mathbf{R}$ represents
the instrumental ``blurring'' of an input spectrum,
and is ideally band diagonal, with a bandwidth that is
neither too large (oversampled) nor too small (undersampled).
All the inferential power
of the spectrum is encapsulated by the ``likelihood functional''
of a model spectrum $\mathbf{m}$ given the data, which
if we assume Gaussian noise can be expressed
by the $\chi^2$ statistic:
\begin{equation}
\label{O09:eq:chispec}
\chi^2_{\mathrm{spec}}(\mathbf{m} ~|~ \mbox{data}) = (\mathbf{f} - \mathbf{R} \mathbf{m})^{\mathrm{T}}
\mathbf{C}^{-1} (\mathbf{f} - \mathbf{R} \mathbf{m}) ~.
\end{equation}

Let us now consider the raw data and calibrations from which
$\mathbf{f}$, $\mathbf{C}$, and $\mathbf{R}$ are derived.
Generally speaking,
the predicted counts $c_i$ in CCD pixel $i$ (where $i$ is a single
index that is understood to run over
rows, columns, and multiple CCDs) may be related to the input model spectrum
$m (\lambda)$ through the linear relation
\begin{equation}
c_i = \int d \lambda \, A_i (\lambda) \, m (\lambda)~,
\end{equation}
where $A_i (\lambda)$ is a transfer function for pixel $i$.
If we assume sufficient finite sampling points of the
spectrum and transfer function in the wavelength domain indexed by
$j$, we may represent the operation of the
telescope, instrument, and detector by a matrix relationship
\begin{equation}
c_i = A_{ij} m_j~.
\end{equation}
Here, the matrix $\mathbf{A}$, which we shall refer to as the
``calibration matrix'', generalizes and incorporates the
spectrum trace solution, wavelength solution, 2D spectrograph
PSF, relative and absolute throughput variation, variations
in CCD pixel sensitivity, and all other calibration considerations
that are generally considered separately from one another.
The problem of accurately estimating $\mathbf{A}$ given
standard calibrations (i.e., arcs and flats, as well as science
frames for self-calibration)
is perhaps the most challenging aspect of spectroscopic data reduction.
The strategy for determining $\mathbf{A}$ is beyond the scope
of this contribution; here we only note that in most
instrumental contexts, the calibration matrix
will have sufficient symmetry, sparsity, and smoothness to
make its estimation a well-constrained problem
\citep{2011pandey_thesis}.

If we assume that $\mathbf{A}$ has been determined, that we have a set
of measured science CCD pixel counts represented by the vector $\mathbf{p}$,
and that the associated (uncorrelated) squared pixel errors constitute
the diagonal raw-pixel covariance matrix $\mathbf{N}$,
the likelihood functional of any model input spectrum vector $\mathbf{m}$
is represented by
\begin{equation}
\label{O09:eq:chiraw}
\chi^2_{\mathrm{raw}}(\mathbf{m} ~|~ \mbox{data}) = (\mathbf{p} - \mathbf{A} \mathbf{m})^{\mathrm{T}}
\mathbf{N}^{-1} (\mathbf{p} - \mathbf{A} \mathbf{m}) ~,
\end{equation}
and in principle any and all inference can be made directly
against the raw pixels.  In practice this approach is generally both
onerous and unnecessary.  However, the obvious analogy between
Equations \ref{O09:eq:chispec} and~\ref{O09:eq:chiraw} suggests
the following operational definitions for three
commonly understood stages of spectral data analysis:
\begin{center}
\begin{tabular}{lcr}
Calibration & $\equiv$ & Likelihood functional determination \\
Extraction & $\equiv$ & Likelihood functional compression \\
Measurement & $\equiv$ & Likelihood functional projection
\end{tabular}
\end{center}
The quality of an extraction algorithm can thus be judged by
the lossiness of its compression of the input-spectrum
likelihood functional through the translation of
$\chi^2_{\mathrm{raw}}$ (as specified by $\mathbf{A}$, $\mathbf{N}$, and $\mathbf{p}$)
into $\chi^2_{\mathrm{spec}}$ (as specified by $\mathbf{R}$, $\mathbf{C}$, and $\mathbf{f}$).

To arrive at a lossless extraction algorithm,
note first that we can approach Equation~\ref{O09:eq:chiraw}
naively to determine a maximum-likelihood
(minimum-$\chi^2$) solution for
the \textit{input} spectrum $\mathbf{m}$ as
\begin{equation}
\mathbf{m}_{\mathrm{m.l.}} = (\mathbf{A}^{\mathrm{T}} \mathbf{N}^{-1} \mathbf{A})^{-1}
(\mathbf{A}^{\mathrm{T}} \mathbf{N}^{-1}) \mathbf{p}~.
\end{equation}
However, this estimator has undesirable properties as an
extracted spectrum.
The square matrix $(\mathbf{A}^{\mathrm{T}} \mathbf{N}^{-1} \mathbf{A})^{-1}$,
being the inverse of
the symmetric banded non-negative matrix
$\mathbf{A}^{\mathrm{T}} \mathbf{N}^{-1} \mathbf{A}$,
will have significant bandwidth and large
positive/negative fluctuations off the diagonal.
Since this is the covariance matrix of the sample elements of
$\mathbf{m}_{\mathrm{m.l.}}$, the estimated spectrum will
exhibit significant ringing.  This is an expected
consequence of the deconvolution implicit in the
solution for $\mathbf{m}_{\mathrm{m.l.}}$.

This deconvolution problem highlights an apparent advantage
of the row-by-row optimal extraction algorithm:
while row-by-row deconvolves the 2D spectrum
image in the spatial direction (on the strong prior assumption
that the input signal is a one-dimensional spectrum distributed
according to a known profile), it does not deconvolve the
instrumental profile in the wavelength direction, and hence
introduces no ringing.  The price, as seen above,
is that this computation is carried out using a mathematically
incorrect model for the projection of photons onto the detector.
One solution in the context of 2D PSF extraction
would be to impose a regularizing prior on
$\mathbf{m}_{\mathrm{m.l.}}$.  However, regularization of the
input spectrum model is the purview of science analysis,
not of data reduction.  If implemented at this stage, it would break
the model of extraction as likelihood functional compression.

To proceed, consider a diagonalization of the matrix
$\mathbf{A}^{\mathrm{T}} \mathbf{N}^{-1} \mathbf{A}$ as
\begin{equation}
\label{O09:eq:cfactor}
\mathbf{A}^{\mathrm{T}} \mathbf{N}^{-1} \mathbf{A}
= \mathbf{R}^{\mathrm{T}} \mathbf{C}^{-1} \mathbf{R} ~,
\end{equation}
where $\mathbf{R}$ is a square matrix
(in contrast to $\mathbf{A}$, which has many more rows than columns)
and $\mathbf{C}^{-1}$ is a diagonal matrix.
This diagonalization need not necessarily be
in the eigen-decomposition sense.
To factorize in the sense described in B\&S,
we first consider the matrix $\mathbf{U}$ whose columns
are the eigenvectors
of $\mathbf{A}^{\mathrm{T}} \mathbf{N}^{-1} \mathbf{A}$,
and the corresponding diagonal matrix $\mathbf{E}$
of eigenvalues.  Since the original matrix is real,
symmetric, and positive definite,
$\mathbf{U}$ is orthogonal, with its inverse
equal to its transpose, and all its eigenvalues are
real and positive.  Now we define
\begin{eqnarray}
\label{O09:eq:rdef}
\mathbf{R} &\equiv& \mathbf{S} \, \mathbf{U} \, \mathbf{E}^{1/2} \mathbf{U}^{\mathrm{T}} \\
\label{O09:eq:cdef}
\mathbf{C} &\equiv& \mathbf{S}^{2} ~,
\end{eqnarray}
where $\mathbf{E}^{1/2}$ is a diagonal matrix whose entries are the
positive square roots of the entries of $\mathbf{E}$, and
$\mathbf{S}$ is a diagonal matrix defined so that $\mathbf{R}$ is normalized
to unity in each row when summed over columns.
(The more intuitive ``flux-conserving''
normalization associated with a sum over rows is less
straightforward to implement.)
With this definition of $\mathbf{R}$ and $\mathbf{C}$,
Equation~\ref{O09:eq:cfactor} is satisfied exactly.
We have diagonalized the (inverse) covariance matrix,
opting not for the orthogonality of $\mathbf{U}$,
but rather for the band-diagonal locality of $\mathbf{R}$,
which derives from its relationship to the matrix
square root of the band diagonal matrix
$\mathbf{A}^{\mathrm{T}} \mathbf{N}^{-1} \mathbf{A}$.
That is to say, the matrix $\mathbf{R}$ by our definition
will ``look like'' a line-spread function in the sense
understood in astronomical spectroscopy.

We note in passing that the definition of $\mathbf{R}$ that we have adopted,
while well motivated, is not unique.
In fact, the motivation for a smoothly varying line spread
function with wavelength may even be worth accepting
a small degree of off-diagonal covariance.
A desire for a flux-conserving definition of $\mathbf{R}$
may also motivate different choices.
As noted by B\&S, cross-talk
between neighboring spectra of a multiobject spectrograph
presents additional complication.  The particular choices
and approximations adopted will necessarily be
driven by instrumental and scientific context.

Finally, we define our extracted spectrum $\mathbf{f}$
as the ``reconvolution'' of our deconvolved extraction
by the resolution matrix $\mathbf{R}$, thereby removing the ringing:
\begin{equation}
\label{O09:eq:fdef}
\mathbf{f} \equiv \mathbf{R} \, \mathbf{m}_{\mathrm{m.l.}}
= \mathbf{R} (\mathbf{A}^{\mathrm{T}} \mathbf{N}^{-1} \mathbf{A})^{-1}
(\mathbf{A}^{\mathrm{T}} \mathbf{N}^{-1}) \mathbf{p}~.
\end{equation}
Now, using the definitions of $\mathbf{R}$, $\mathbf{C}$, and $\mathbf{f}$
given in Equations~\ref{O09:eq:rdef}--\ref{O09:eq:fdef},
we find that $\chi^2_{\mathrm{spec}}$ as given
by Equation~\ref{O09:eq:chispec} is equal up to a constant
offset to $\chi^2_{\mathrm{raw}}$ as given by
Equation~\ref{O09:eq:chiraw} \textit{for any trial
model spectrum} $\mathbf{m}$.
Thus we attain the
goal of extraction as lossless likelihood-functional
compression,
with $\mathbf{A}$, $\mathbf{N}$, and $\mathbf{p}$ replaced
respectively
by the band-diagonal LSF matrix $\mathbf{R}$,
the diagonal spectrum covariance matrix $\mathbf{C}$
(i.e., an error vector), and the flux
vector $\mathbf{f}$.  Equally important is the fact that
$\mathbf{R}$, $\mathbf{C}$, and $\mathbf{f}$ collectively
satisfy the intuitive understanding
of ``a spectrum'' in the astronomer's sense!

\acknowledgments
The authors thank David Hogg, Sam Roweis, Michael Blanton,
Julian Borrill, and Ted Kisner for valuable discussions of this subject.


\end{document}